\documentclass[a4paper,11pt]{article}
\usepackage{amsmath,epsfig,graphicx,amssymb}
\newcommand{\grad}{{\rm grad}}
\newcommand{\divz}{{\rm div}}

\begin{document}

\begin{center}
{\Large\bf
Thermomagnetic Convection of Magnetic Fluids in a Cylindrical Geometry}\\[.5cm]
Adrian Lange\\[.5cm]
Institut f\"ur Theoretische Physik, Universit\"at Magdeburg, Postfach 4120, D-39016 Magdeburg, Germany\\
\end{center}

\vspace{.2cm}

\begin{abstract}
       The thermomagnetic convection of magnetic fluids in a cylindrical geometry subjected
       to a homogeneous magnetic field is studied. The study is motivated by a novel
       thermal instability [W. Luo {\it et al}., Phys. Rev. Lett. {\bf 82}, 4134 (1999)].
       As model system a composite cylinder with inner heating is considered which reflects
       the symmetry of the experimentally setup. The general condition for the existence of
       a potentially unstable stratification in the magnetic fluid is derived. Within a linear
       stability analysis the critical external induction for the onset of thermomagnetic convection
       is determined for dilute and nondilute magnetic fluids. The difference between both thresholds
       allows to test experimentally whether a test sample is a dilute fluid or not.
\end{abstract}

\begin{tabular}{ll}
{\bf Contact author:}& Adrian Lange, Fax:  +49-391-6711205\\
 & \makeatletter email: adrian.lange@physik.uni-magdeburg.de \makeatother
\end{tabular}

\newpage

\section{Introduction}
Magnetic fluids (MFs) are superparamagnetic fluids formed by a stable colloidal suspension
of ferrimagnetic nanoparticles dispersed in a carrier
liquid  \cite{rosensweig}. The behaviour of MFs is characterized by
the complex interaction of their hydrodynamic and magnetic properties with external forces.
For the phenomenon of thermomagnetic convection these forces are a temperature gradient and a
magnetic field. Most studies consider the geometry of a horizontal
layer which is simultaneously subjected to a vertical temperature gradient and either to
a constant vertical magnetic field \cite{finlayson,schwab,auernhammer00} or to a vertical
magnetic field with a constant gradient \cite{huang}. Studies for a cylindrical geometry
are few \cite{morimoto98,zebib96} and focus on the thermomagnetic convection under
microgravity \cite{odenbach93}.

The present analysis of thermomagnetic convection in a cylindrical geometry is motivated
by a recently observed novel convective instability.
In \cite{du98,luo99} a horizontal layer of MF  between two glass plates is locally
heated by a focused laser beam. It passes perpendicularly through
the layer in the presence of a homogeneous vertical magnetic
field. The absorption of the light by the fluid generates a temperature
gradient and subsequently a refractive index gradient. This gradient is
optically equivalent to a diverging lens, leading to an enhancement of the
beam divergence. As result, a stationary diffraction pattern of concentric rings 
is observed for zero magnetic field. Above a certain threshold of the magnetic
field, the circular rings are replaced by polygonally shaped patterns which switch
among different shapes alternatively. Based on {\it numerics} it was stated that the
characteristic time scales for mass  and thermal diffusion are equal \cite{du98}.
Thus thermal conduction and thermal diffusion contribute equally to the formation
of the diverging lens. The polygonally shaped diffraction patterns were interpreted
as `fingerprints' of vertical convection columns \cite{luo99}.

Both claims are controversial due to the following reasons. An analysis of the {\it physical
quantities} reveals that both characteristic time scales are several orders of magnitude
apart (below and \cite{shliomis00}). Direct spatial temperature
measurements \cite{schaertl99} and independent measurements of all relevant material
parameters for organic dispersions \cite{spill00} showed that the ring pattern
is essentially caused by the temperature contribution and not by the concentration
contribution. The determination of the relevant time scales confirmed that the
experiment is dominated by the characteristic time for convection. Presently there is
no sound theoretical description for the dependence of the size of the outmost of
the concentric rings on the laser power (see Fig.~2 in \cite{spill00}).

A major hindrance of the studied system in \cite{du98,luo99} is that it is almost impossible
to gain information about the spatial distribution of temperature and concentration inside
the MF layer. Due to the lack of sound internal information many hypotheses can be
brought forward to explain the experimental results. Among them is the recent discussion
whether the shape instability of a hot nonmagnetic bubble surrounded by MF can
be accounted for the observed phenomena \cite{shliomis_01,lu01}.

This situation motivates the present work in which a model system is studied which
reflects the essentials of the experimental setup. These are the axis-symmetry of the
heating and the finite hight and width of the layer. The aim is to determine the necessary
conditions and the critical magnetic inductions for the appearance of axial convection columns.
Thus the focus is on the general requirements for thermomagnetic convection in an
axis-symmetric heated system.

This paper is organized as follows: The system and the relevant equations of the problem
as well as the condition for a potentially unstable stratification in the fluid are displayed
in the next section. Based on  a linear stability analysis (Sec.\ III), the results are
presented and discussed in Sec.\ IV. In the final section, the results are summarized.

\section{Model and Equations}
The model system is given by a composite, circular cylinder of height $h$ which consists
of three parts. The inner cylinder of radius $R_1$, constant temperature $T_1$, and constant
susceptibility $\chi(T_1)$ is surrounded by a middle cylinder of radius $R_2$.
In the gap $R_2-R_1$ the temperature decreases to $T_0<T_1$ and consequently $\chi$ is a
spatially varying quantity, $\chi=\chi[T(r)]$. The outer cylinder
has the radius $R_{{\rm out}}$, where in the region $R_{{\rm out}}-R_2$ the constant temperature
$T_0$ and constant susceptibility $\chi(T_0)$ is present. The whole system is subjected to a
homogeneous vertical magnetic field and its effective susceptibility is
given by
\begin{equation}
\label{eq:1}
\chi_{{\rm eff}}={1\over R_{{\rm out}}}\biggr( R_1\chi(T_1) +(R_2-R_1)\int_{R_1}^{R_2}\!\!\! dr\,
\chi[T(r)]+(R_{{\rm out}}-R_2)\chi(T_0)\biggr)\; .
\end{equation}

In the presence of a uniform external magnetic induction ${\bf B}_{{\rm ext}}$, the internal field
in the gap is given by ${\bf H}_{{\rm int}}={\bf B}_{{\rm ext}}/[\mu_0(1+N\chi)]$. The susceptibility
of the MF is  $\chi =\chi_L (1+\beta_1 \chi_L)$, where $\chi_L$ is the susceptibility
according to Langevins theory which assumes non-interacting particles.  Higher order terms in
$\chi_L$ are included in order to determine the magnetic (or Kelvin) force density
beyond the dilute limit $\chi=\chi_L$. The coefficient $\beta_1$ was determined
in different microscopic models \cite{onsager36,mean_spheric,buyevich92} which all
provide the same value $\beta_1=1/3$. The demagnetization factor $N$ accounts for the finite
size of the composite cylinder and is a function of the heigh-to-diameter ratio
$\gamma=h/(2R_{{\rm out}})$ and the effective susceptibility $\chi_{{\rm eff}}$ \cite{chen91}.
The Kelvin force follows then as \cite{lange01_kelvin}
\begin{equation}
\label{eq:2}
{\bf f}_{{\rm K}} =-{B_{{\rm ext}}^2\over \mu_0}F_{\chi_L}{\grad\chi_L\over \chi_L}\, ,
\end{equation}
where
\begin{equation}
\label{eq:3}
F_{\chi_L}={\chi_L^2\left\{N+\beta_1\left[3N\chi_L\left( 1+
\beta_1\chi_L\right)-1\right]\right\}\over (1+N\chi)^3}\; .
\end{equation}

Considering MFs as binary mixtures, it is necessary to evaluate
the influence of temperature and concentration on pattern phenomena by analyzing the
corresponding time scales. These are the characteristic time for convection
$t_c=L_c^2/\kappa$ and for mass diffusion $t_d=L_d^2/D$, where $L_c$ ($L_d$) is the
typical length for convection (diffusion), $\kappa$ the thermal diffusivity, and $D$
the mass diffusion coefficient. Using the data given in \cite{rosensweig,du98}, one gets
$\kappa\sim 4\times 10^{-8}$ m$^2\,$s$^{-1}$ and $D\sim 8\times 10^{-12}$ m$^2\,$s$^{-1}$.
Except in special designed geometries as in \cite{voelker}, where $L_c\ggg L_d$,
the characteristic time for diffusion is of three orders of magnitude larger than the
characteristic time for convection. Since in our model both length scales are equal
to the gap width $R_2-R_1$, diffusion phenomena can be neglected.

The system is governed by the equation of continuity, the Navier-Stokes equations,
and the equation of heat conduction for the MF which are in nondimensional form 
\begin{eqnarray}
\label{eq:4}
\divz\,\bar{\bf v} &=&0 \, ,\\
\label{eq:5}
{\partial\bar{\bf v}\over \partial \bar t}+(\bar{\bf v}\grad )\bar{\bf v} &=&
{\cal P}\left( -\grad\,\bar p +\Delta\bar{\bf v}\right)
+{\cal M}F_{\chi_L}{\grad\bar T\over \bar T} \, ,\\
\label{eq:6}
{\partial \bar T\over \partial \bar t}+(\bar{\bf v}\grad )\bar T &=&\Delta \bar T\, ,
\end{eqnarray}
where the Prandtl number ${\cal P}=\nu/\kappa$ characterizes the fluid and
the magnetization number ${\cal M}=B_{{\rm ext}}^2 (R_2-R_1)^2/(\mu_0\rho\kappa^2)$
tunes the external excitation. Denoting $\nu$ as kinematic viscosity, the
velocity ${\bf v}=(u,v)$ is scaled with $\kappa/(R_2-R_1)$, time with
$(R_2-R_1)^2/\kappa$, temperature with $(T_1-T_0)$, and pressure $p$ with
$\rho\kappa\nu/(R_2-R_1)^2$. $\Delta$ and $\grad$ are the corresponding differential
operators in the plane cylindrical coordinates $\bar r$ and $\phi$. Rigid boundary
conditions are assumed for the velocity at the inner and outer radius of the gap,
$\bar u=\partial_{\bar r}\bar u=0$ at $\bar r=\eta/(1-\eta)$ and $\bar r=1/(1-\eta)$,
where the radii ratio is given by $\eta=R_1/R_2$. The temperature is assumed to
be constant at each boundary, $\bar T\bigr(\bar r=\eta/(1-\eta)\bigr)=\bar T_1$ and
$\bar T\bigr(\bar r=1/(1-\eta)\bigr)=\bar T_0$.

\begin{figure}[htbp]
  \begin{center}
    \includegraphics[height=4.0cm,width=8.25cm]{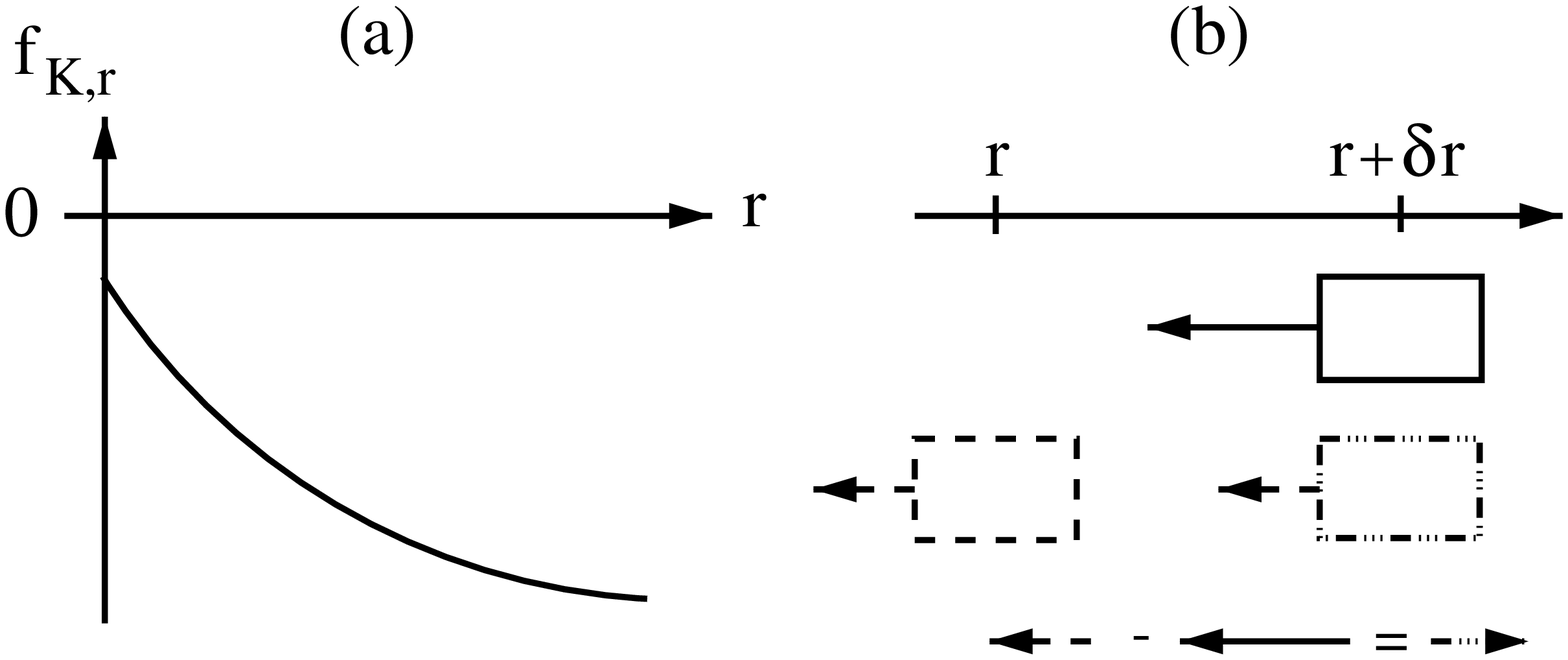}
    \caption{Lange, Physics of Fluids}
    \label{fig:1}
  \end{center}
\end{figure}

Since the Kelvin force is the only destabilizing force present in the system,
one has to determine which profile leads to a potentially unstable stratification
in the fluid. For heating at the inner radius, the required profile is given in
Fig.~\ref{fig:1}(a): the $r$-component of the Kelvin force density has to act
inwards and its absolute value increases monotonically outward.
With such a profile a fluid volume at the distance $r+\delta r$ (solid rectangle in
Fig.~\ref{fig:1}(b)) experiences a larger force towards the center compared to
a fluid volume at the distance $r$ (dashed rectangle). Moving the latter fluid
volume from $r$ to $r+\delta r$ (dot-dashed rectangle) results in an effective
force which points in the direction of the displacement (indicated symbolically
by the subtraction of the arrows in Fig.~\ref{fig:1}(b) bottom). This force may
enhance small displacements of warmer fluid volumes towards cooler regions and
thus making the stratification potentially unstable.

\begin{figure}[htbp]
  \begin{center}
    \includegraphics[height=6.0cm,width=8.5cm]{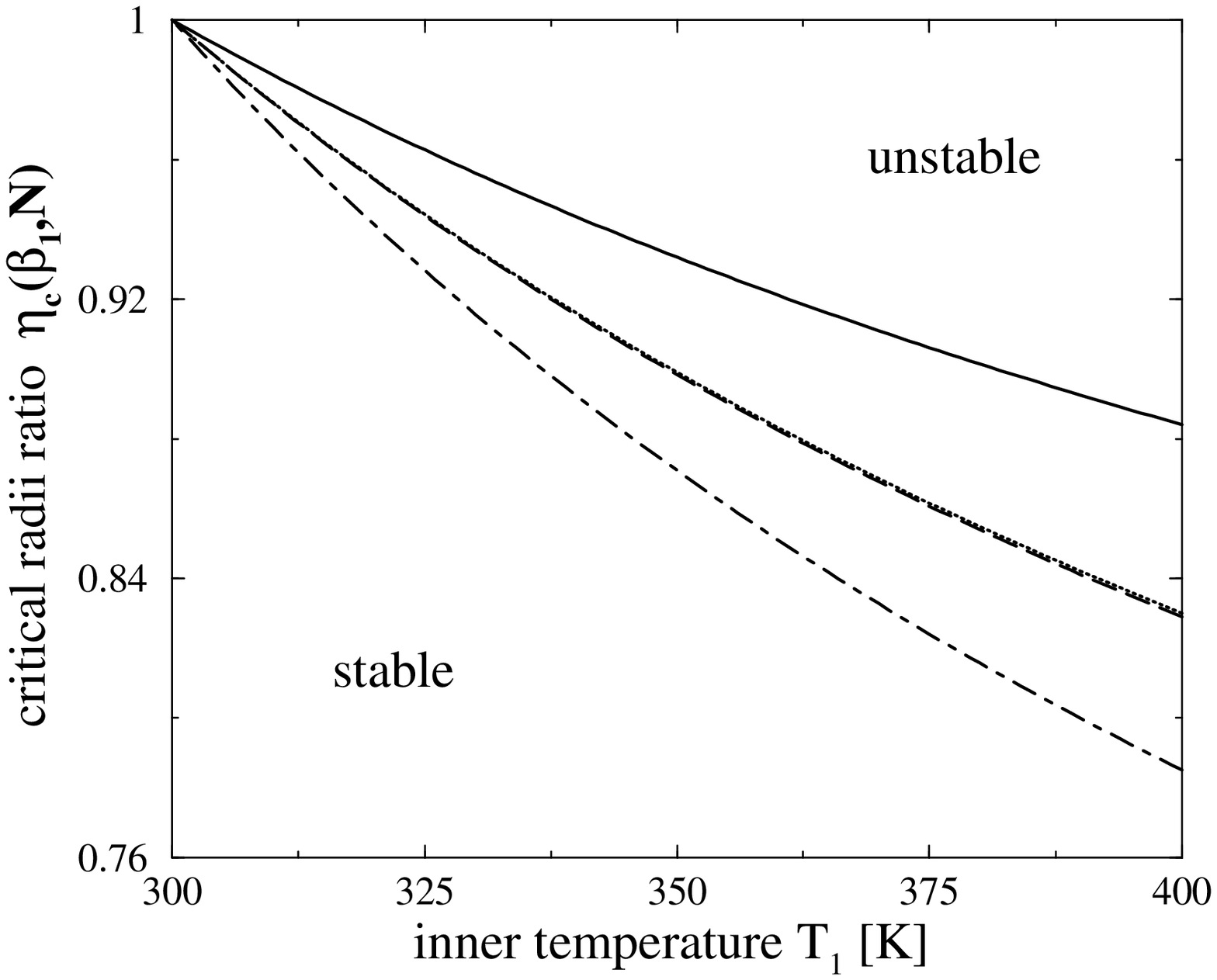}
    \caption{Lange, Physics of Fluids}
    \label{fig:2}
  \end{center}
\end{figure}

The above argument has to be tested for the quiescent conductive state which is
given by $\bar{\bf v}_G =0$ and
$\bar T_G = \bar T_0 +(\bar T_1-\bar T_0)\ln [\bar r(1-\eta)]/\ln\eta$. Applying the
condition for a destabilizing force profile to the $r$-component of the Kelvin
force density in Eq.~(\ref{eq:2}) leads to the condition
\begin{eqnarray}
\nonumber
{\partial\over \partial\bar r}{\rm f}_{{\rm K},\bar r} &=&
  -{\cal M}\left( {\partial_{\bar r} \bar T\over \bar T}\right)^2
   {\partial (\chi_L F_{\chi_L})\over \partial\chi_L}
  +{\cal M} F_{\chi_L} {\partial_{\bar r}^2 \bar T\over \bar T} < 0\\
\label{eq:7}
&&{\rm for~all~~} \bar r\in[\eta/(1-\eta), 1/(1-\eta)] \; .
\end{eqnarray}
Since $\partial_{\bar r}{\rm f}_{{\rm K},\bar r}$ is a monotonously decreasing function
of $\bar r$, it is sufficient if ~$\partial_{\bar r}{\rm f}_{{\rm K},\bar r}<0$~ at
~$\bar r=\eta/(1-\eta)$ in order to fulfil the condition~(\ref{eq:7}). Depending on the
temperatures $\bar T_1$ and $\bar T_0$, the
Langevin susceptibility $\chi_L$, the demagnetization factor $N$, and $\beta_1$ the
condition~(\ref{eq:7}) entails that the radii ratio $\eta$ has to be larger than a
critical value. For realistic temperatures $T_1$ above a room
temperature of $T_0=300$ K, it becomes clear that this condition is met only
in a narrow gap (see Fig.~\ref{fig:2}). This is plausible because the nonlinear 
temperature profile $\bar T_G\sim\ln [\bar r(1-\eta)]$ can be well approximated in a small
gap by a linear profile which always satisfies the requirement (\ref{eq:7}) if
$\beta_1 =0$.

\section{Linear Stability Analysis}
Exploiting the smallness of the gap, in the linear stability analysis terms as
$\partial_{\bar r} (\partial_{\bar r}+1/\bar r)$ are approximated by
$\partial_{\bar r}^2$ and the new variable $\zeta=\bar r-\eta/(1-\eta)$ is
introduced. All small disturbances from the ground state are decomposed into normal
modes, i.e. into components of the form
$[\bar u, \bar p, \bar T] ={\rm e}^{n\bar t}\cos(l\phi)[\bar u(\zeta), \bar p(\zeta), \bar T(\zeta)]$
and $\bar v ={\rm e}^{n\bar t}\sin(l\phi)\bar v(\zeta)$, respectively. The
nondimensional growth rate is denoted by $n$ and $l$ is the azimuthal wave number.
For marginal stability, $n\equiv 0$, the differential equations to solve are
\begin{eqnarray}
\label{eq:8}
\left({\partial^2 \over \partial\zeta^2}-\alpha^2\right)^2\!\!\!\!\bar u
-{\alpha^2\over l^2}\!\left({\partial^2 \over \partial\zeta^2}-\alpha^2\right)\!\! \bar u
&=&-\alpha^2{{\cal M}\over {\cal P}} f_{\chi_L} {\bar T\over \bar T_G^2}
{\partial \bar T_G\over \partial\zeta}\, ,\\
\label{eq:9}
\left( {\partial^2\over \partial\zeta^2} -\alpha^2\right)\! \bar T &=& \bar u
\, (\bar T_0-\bar T_1)\, ,
\end{eqnarray}
where $\alpha =(1-\eta)l/\eta$ ~and
\begin{eqnarray}
\nonumber
f_{\chi_L}&=&-\chi_L{\partial F_{\chi_L} \over \partial\chi_L}
  = {\chi_L^2 \over (1+N\chi)^4}[ 6 N^2\beta_1^2\chi_L^3(1+\beta_1\chi_L)\\
\label{eq:10}
&&+4 N\chi_L^2\beta_1 (N-4\beta_1 )+\chi_L(N-10\beta_1 )+2\beta_1-2 N]\; .
\end{eqnarray}
In order to satisfy the four boundary conditions $\bar u=\partial_\zeta \bar u=0$ at
$\zeta=0, 1$, the ansatz
\begin{equation}
\label{eq:11}
\bar u(\zeta)=\sum_{m=1}^K P_m\Bigr[ \sinh(a\zeta)+B_m\cosh(a\zeta)+C_m\sin(b\zeta)
+D_m\cos(b\zeta)\Bigr]
\end{equation}
with $a=\sqrt{q_m^2+\alpha^2}$ and $b=\sqrt{q_m^2-\alpha^2}$ is chosen. $q_m$ is
a root of a transcendental equation and the constants $B_m$, $C_m$, and $D_m$
are determined by the boundary conditions (for details see \cite{chandrasekhar}). 
With Eq.~(\ref{eq:11}) the solution of Eq.~(\ref{eq:9}) reads
$\bar T(\zeta)= \sum_{m=1}^K P_m\left[ T_m(\zeta)+C_1{\rm e}^{\alpha\zeta}+
C_2{\rm e}^{-\alpha\zeta}\right]$. $T_m(\zeta)$ is the solution of the inhomogeneous
equation (due to its lengths not given here) and the constants $C_1$ and $C_2$ are
determined by the boundary conditions $\bar T=0$ at $\zeta=0, 1$. Using
Eq.~(\ref{eq:11}) and the solution for $\bar T(\zeta )$, Eq.~(\ref{eq:8}) can be
approximately solved by the Galerkin method. Due to the good convergence, all presented
results are based on the third approximation (see Table~\ref{table1}). For
the calculations fluid parameters of EMG 901 are used: $\rho=1.53\times 10^3$ kg$\,$m$^{-3}$,
$\nu=6.54\times 10^{-6}$ m$^2\,$s$^{-1}$, $\chi_L=3$, and
$\kappa =4.2\times 10^{-8}$ m$^2\,$s$^{-1}$ \cite{comment}. The temperature
at the outer radius of the gap $R_2=1$ cm is fixed at $T_0=300$ K. The hight of the
composite cylinder is given by $h=1$ cm and the inner radius by
$\eta=1.01\eta_c(\beta_1=1/3, N=1)$. The choice of $\eta_c(\beta_1=1/3, N=1)$
(solid line in Fig.~\ref{fig:2}) ensures that for all following parameter sets
the condition~(\ref{eq:7}) is fulfilled. 

\section{Results and Discussion}
\begin{figure}[htbp]
  \begin{center}
    \includegraphics[height=6.0cm,width=8.5cm]{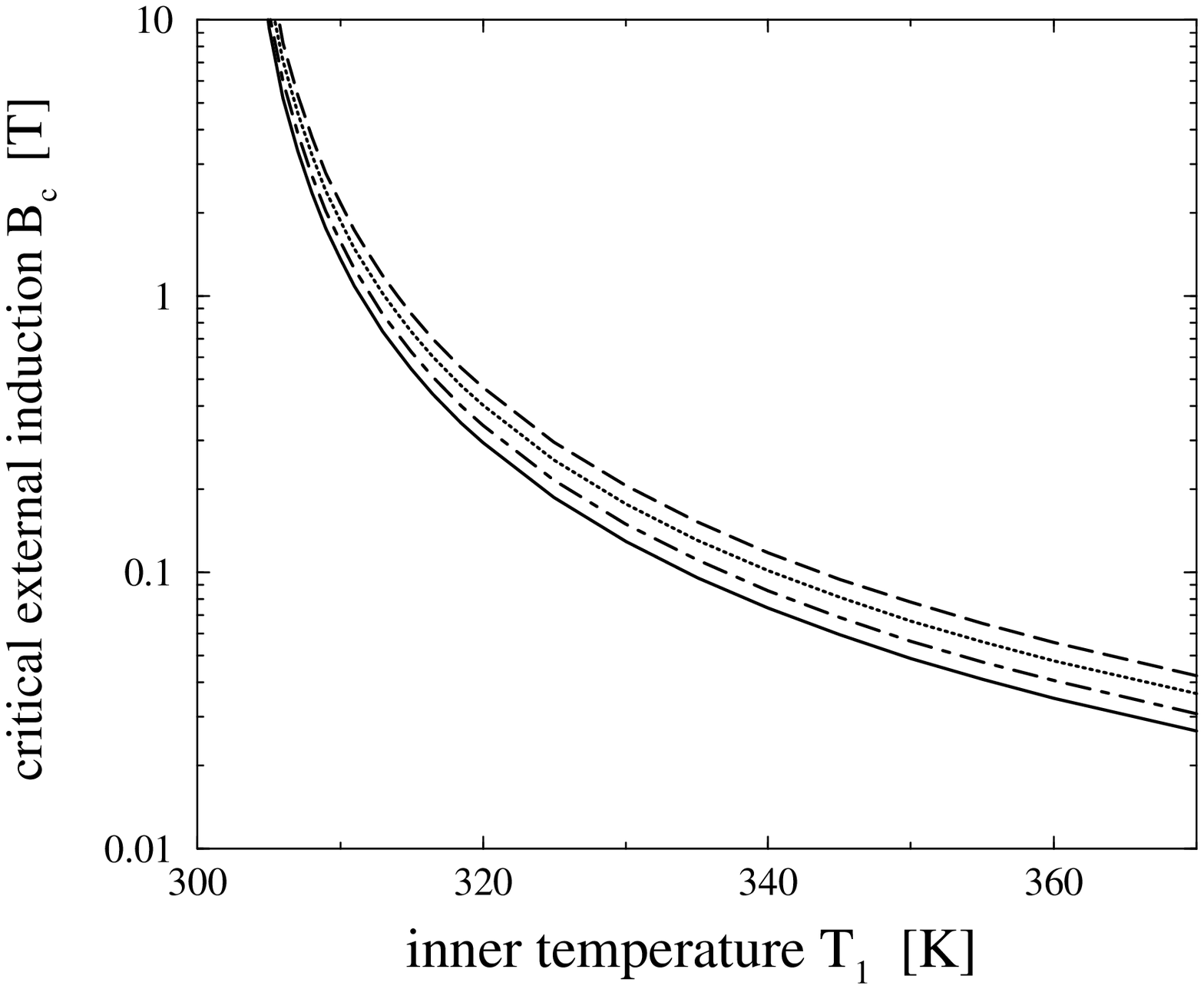}
    \caption{Lange, Physics of Fluids}
    \label{fig:3}
  \end{center}
\end{figure}

\begin{figure}[htbp]
  \begin{center}
    \includegraphics[height=6.0cm,width=8.5cm]{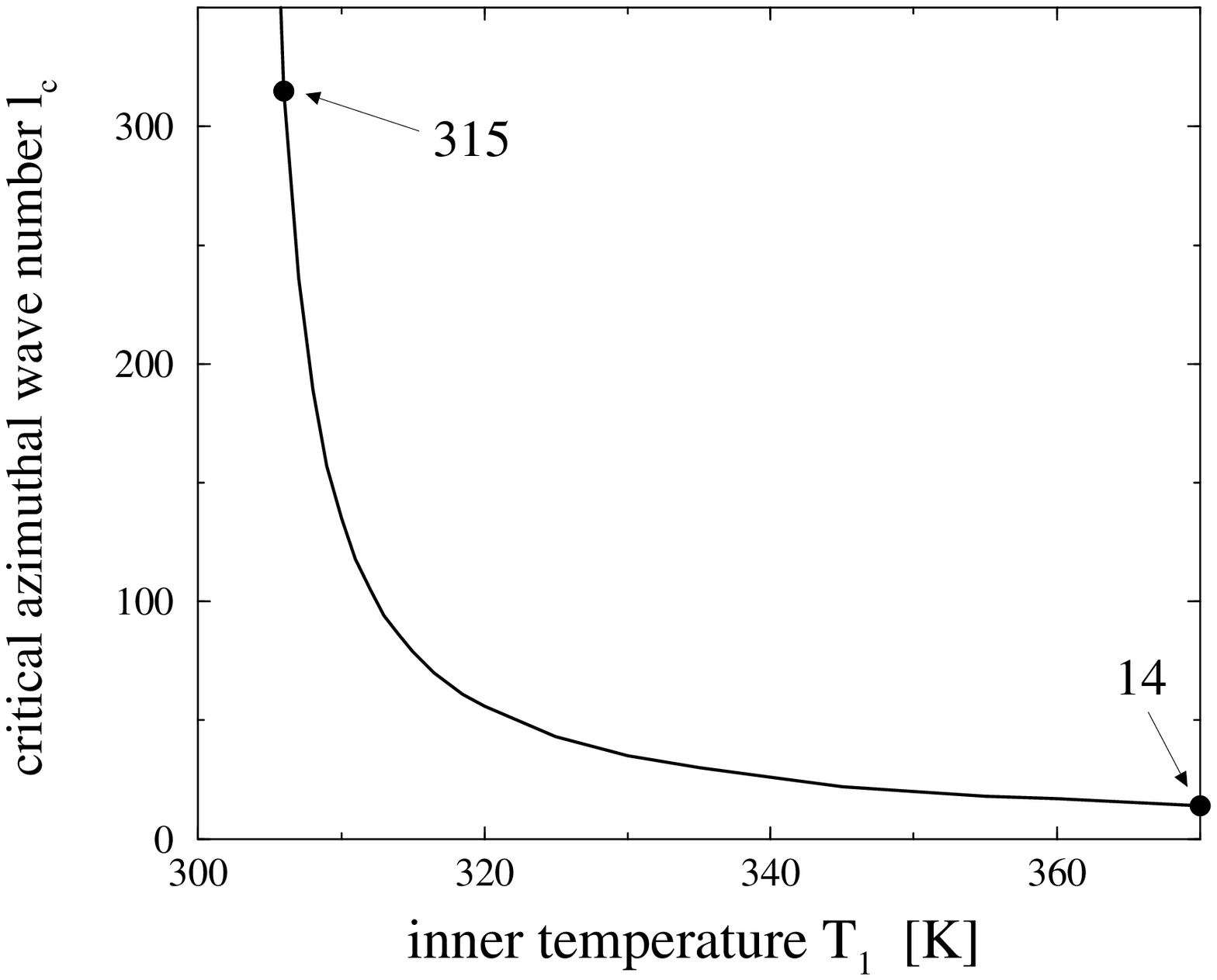}
    \caption{Lange, Physics of Fluids}
    \label{fig:4}
  \end{center}
\end{figure}

Solving Eq.~(\ref{eq:8}) with the Galerkin method and subsequently minimization
with respect to the azimuthal wave number determines the critical external
induction $B_c$ and the corresponding wave number $l_c$ (Figs.~\ref{fig:3} and \ref{fig:4}).
Four different parameter sets were chosen: a dilute ($\beta_1=0$) and a nondilute
($\beta_1=1/3$) MF with $R_{{\rm out}}=\infty$ ($N=1$) and $R_{{\rm out}}\simeq 3.33$
cm, respectively. The demagnetization factor $N$ for the resulting
height-to-diameter ratio $\gamma =0.15$ and the effective susceptibility $\chi_{{\rm eff}}$
of the composite cylinder accordingly to Eq.~(\ref{eq:1}) is taken from the
data given in \cite{chen91}.

Decreasing the temperature difference from $\Delta T=70$ K to $\Delta T=4$ K
causes a dramatic increase in the critical induction of nearly three orders
of magnitude (Fig.~\ref{fig:3}). With decreasing temperature difference the critical radii ratio
grows, i.e. the allowed gap becomes more narrow. Since the convection rolls prefer
the same length scale in $r$ and $\phi$-direction, much more rolls have to be
driven in a very small gap. The energy for this effort comes from the external
induction which is why it amplifies drastically for small $\Delta T$
(Fig.~\ref{fig:3}).

Whereas the critical azimuthal wave number $l_c$ is independent of $\chi(\chi_L )$
and $N$ (Fig.~\ref{fig:4}), the critical induction varies. First the two
thresholds for the case of an infinitely extended layer, $R_{{\rm out}}=\infty$,
are compared. The inclusion of a quadratic
term in the susceptibility with $\beta_1=1/3$ results in a
lower threshold for the onset of convection than in the dilute case
$\beta_1=0$ (solid and long-dashed line in Fig.~\ref{fig:3}). The difference between
the thresholds is nearly the same value,
$B_c (\beta_1\!\!=\!\!1/3, N\!\!=\!\!1)\simeq 0.63 B_c (\beta_1\!\!=\!\!0, N\!\!=\!\!1)$,
for all tested temperatures $304$ K $\leq T_1\leq$ $370$ K.

Now the thresholds for the case of a finite layer with  $R_{{\rm out}}\simeq 3.33$ cm are
compared. Contrary to the previous case, the threshold for a dilute fluid
(dot-dashed line in Fig.~\ref{fig:3}) is lower than for a nondilute fluid (dotted line).
Again the difference is almost constant over the entire temperature range,
$B_c [\beta_1\!\!=\!\!1/3, N(\gamma\!\!=\!\!0.15,\chi_{{\rm eff}})]
\simeq 1.18 B_c [\beta_1\!\!=\!\!0, N(\gamma\!\!=\!\!0.15,\chi_{{\rm eff}})]$.

\begin{figure}[htbp]
  \begin{center}
    \includegraphics[height=6.0cm,width=8.5cm]{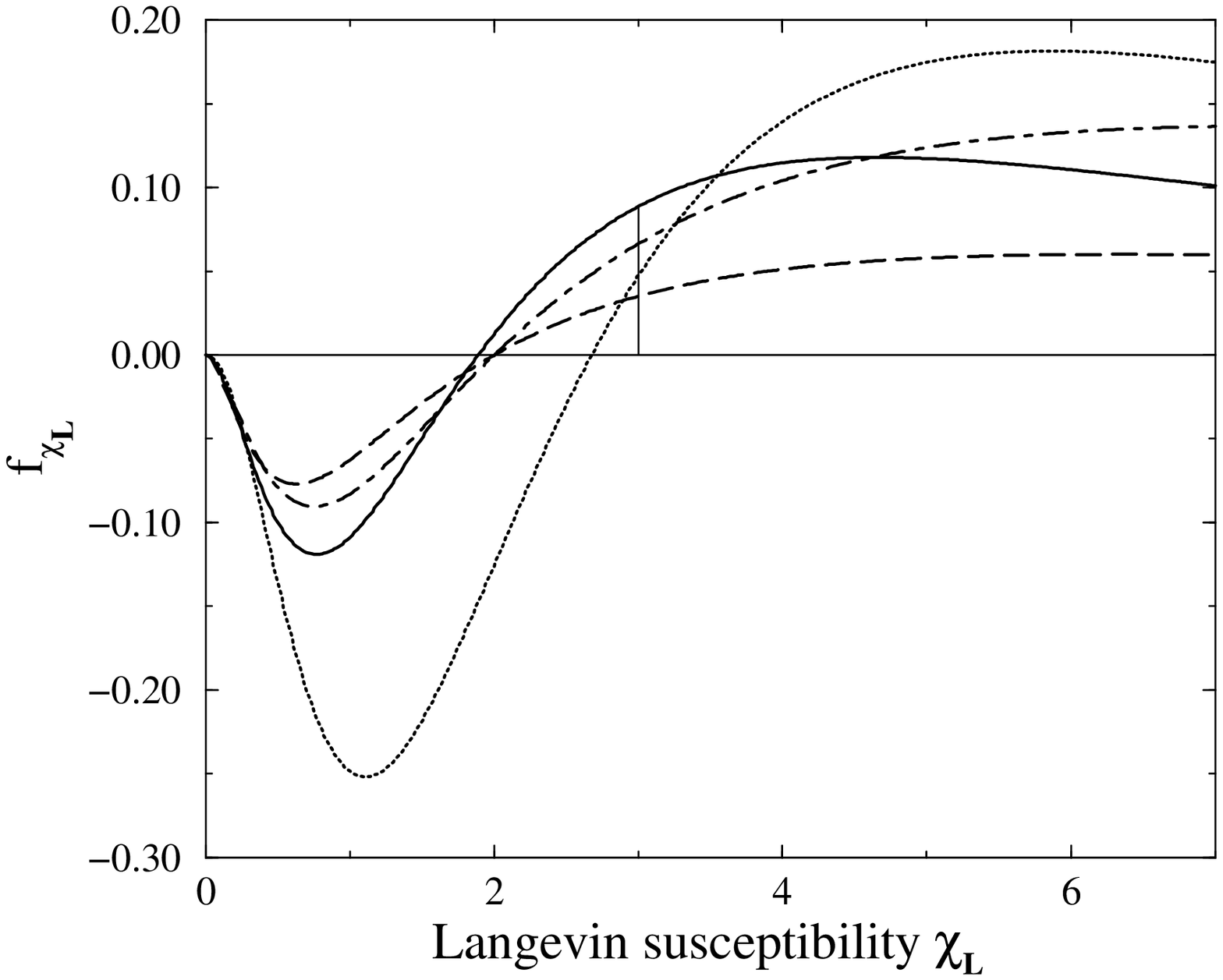}
    \caption{Lange, Physics of Fluids}
    \label{fig:5}
  \end{center}
\end{figure}

The relation of the different thresholds is caused by the value of $f_{\chi_L}$
for the given combinations of $N$, $\beta_1$, and $\chi_L$. $f_{\chi_L}$ can be
considered as a measure for the strength of the magnetic force in the gap: as higher the
value of $f_{\chi_L}$ as lower the critical external induction necessary to trigger
the convection. Figure~\ref{fig:5} shows the value of $f_{\chi_L}$ for the four
considered parameter sets. At $\chi_L=3$ the relation $f_{\chi_L}(\beta_1\!\!=\!\!0,N\!\!=\!\!1) <
f_{\chi_L}(\beta_1\!\!=\!\!1/3,N\!\!=\!\!1)$
(see cross-sections of the long-dashed and the solid line with the vertical solid line)
is the reason that the threshold for the dilute fluid is higher than for the nondilute fluid.
The opposite relation $f_{\chi_L}[\beta_1\!\!=\!\!0,N(\gamma\!\!=\!\!0.15,\chi_{{\rm eff}})] >
f_{\chi_L}[\beta_1\!\!=\!\!1/3,N(\gamma\!\!=\!\!0.15,\chi_{{\rm eff}})]$
(see cross-sections of the dot-dashed and the dotted line with the vertical solid line)
causes the opposite relation for the thresholds in the case of a finite layer.

The physical reasons which cause theses differences are the following. The Kelvin
force is proportional to the magnetization in the magnetic fluid. Thus as higher the
magnetization is, as lower the external induction can be in order to generate the same
strength of the magnetic force. In the infinite case, where $N=1$ is independent of $\chi$,
a higher concentration of magnetic particles in the fluid
leads to a higher magnetization and therefore to a lower threshold. In the finite case,
the demagnetization factor depends on $\chi$ \cite{chen91} and is smaller for higher
concentrations of magnetic particles than for lower concentrations. A higher demagnetization
factor means a lower inner field and a higher magnetization, respectively. Therefore in
the finite case an increase in the concentration results in two counteracting effects
with respect to the magnetization. In the studied example of $\chi_L=3$ the influence
of the demagnetization effect wins: the dilute fluid has the lower threshold. But for
$\chi_L>3.2$ the direct influence of the concentration succeeds over the demagnetization
effect. The nondilute fluid has the lower threshold (the dotted line is then above the
dot-dashed line, see Fig.~\ref{fig:5}).

In the infinite and finite case the clear and measurable difference between the thresholds
opens a very good opportunity to decide whether a test sample is a dilute fluid or not. Just
by measuring the threshold for the onset of convection in the proposed model system the answer
can be given. The critical induction depends on the fluid and system parameters as
$B_c\sim \kappa\sqrt{\rho }/(R_2-R_1)$. By choosing fluids with low (high) density and thermal
conductivity and a large (small) radius $R_2$, the threshold can be lowered (raised) corresponding
to the experimentally available magnetic fields.

The major obstacle in order to compare the results with the experimental data in \cite{luo99}
is the lack of an experimentally determined spatial profile of the temperature inside the sample.
Therefore it is not possible to extract an estimation what might be the values of $R_1$ and $R_2$
in the experiment. Nevertheless the calculated values indicate that really high critical external
inductions are necessary to trigger vertical convections rolls by a {\it pure  radial} temperature
gradient. The threshold for the induction reaches extremely high values if one extrapolates
towards radii in the range of hundreds of micrometers, not unlikely due to the focused laser beam
used in the experiment \cite{du98,luo99}. This leads to the conclusion that vertical convections rolls
due to a pure radial temperature gradient are unlikely to account for the observed phenomena of
polygonally shaped diffraction patterns.

Due to the lack of information from inside the sample, it is not clear whether the temperature
profile in the experimental sample is purely radial. There are hints in \cite{luo99_jmmm}
that due to the absorption along the way of the laser beam a vertical temperature distribution
exists as well. A further cause for an axial temperature gradient is the heat loss through the
glass plates by which the MF layer is sandwiched. If such a vertical temperature gradient
comes into play, concentration gradients due to the Soret effect may become important. The
relative influence of temperature and concentration gradients is strongly effected by the relation
of the characteristic times $t_c$ and $t_d$ which depend quadratically on the lengths $L_c$ and $L_d$,
respectively. With a radial and a vertical temperature gradient present, it becomes even more
important to have reliable data of the internal profiles to estimate these lengths.

Two differences between the model and the motivating experiment should be noted. The constant
temperature $T_1$ for the inner cylinder is not given in the experiment. How much the temperature
varies in this inner area is not known. The numerical calculations in \cite{du98} suggest a
difference of about $15$ K. The thresholds were calculated for rigid boundaries whereas in the
experiment the fluid layer boundary is free. For thermal convection in a rotating layer of magnetic fluid
the influence of rigid and free boundaries on the threshold was studied in \cite{auernhammer00}.
Considering the case of zero rotation, the thresholds differ by not more than $20$\% (see
Figs.~3 and 4 in \cite{auernhammer00} for rotation number $T\rightarrow 0$). With respect to
the above mentioned major obstacle, these differences may alter the results only marginally.

\section{Summary}
A model system of a composite cylinder of finite size with axis-symmetrical temperature
distribution is presented. Using the Kelvin force density~(\ref{eq:2},\ref{eq:3}) a
general condition~(\ref{eq:7}) is derived for which a potentially unstable stratification
exists if the inner cylinder is heated. Depending on the temperature difference, the size of
the composite cylinder and the dilute or nondilute character of the magnetic fluid,
the critical gap sizes are calculated. The general result is that only in a narrow gap the
requirement for a potentially unstable stratification is met. Exploiting this property, a linear
stability analysis is performed in order to determine the critical external induction
for the onset of thermomagnetic convection. With decreasing temperature difference,
the critical induction increases dramatically. The reason is that for smaller temperature
differences which demand smaller gaps in order to fulfil condition~(\ref{eq:7}),
much more convection rolls have to be driven. The driving of these many rolls causes
the drastic increasing of the critical induction. The distinct difference between
the threshold for dilute and nondilute magnetic fluids allows to use the considered
system for an experimental determination whether a test fluid is a dilute or nondilute
one.

The rather high external induction, needed to stimulate the convection flow, leads to the
conclusion that vertical convections rolls due to a pure radial temperature gradient are unlikely to 
account for the observed diffraction patterns. The consideration of a vertical temperature
gradient entails that concentration gradients may become relevant. To answer this question
the characteristic time for the diffusion with respect to whose for convection has to
be re-estimated. Because they will not be necessarily apart by orders of magnitude as in the
case of a pure radial temperature gradient. Also different convection patterns can be expected
with the presence of a vertical temperature gradient. In order to come to a correct statement
about the contribution of mass and thermal diffusion to the diffraction patterns in magnetic
fluids, spatial temperature measurements and independent measurements of all relevant material
parameters as in \cite{schaertl99,spill00} for an organic dispersion are highly desirable.

\vfill\eject
\begin{table}
\caption{Critical external induction $B_c$ in dependence of the order of
approximation for $N=1$.}
\label{table1}
\begin{tabular}{cccccc}
\noalign{\smallskip}
$T_1 [K]$ & $\beta_1$ &\multicolumn{4}{c}{$B_c$ [T]}\\
          &         &  K=1    & K=2    & K=3    & K=4 \\
\noalign{\smallskip}\hline\noalign{\smallskip}
306    &   0     & 8.2527   & 8.2527  & 8.2502  & 8.2502  \\
       &   1/3   & 5.1952   & 5.1952  & 5.1937  & 5.1937  \\
320    &   0     & 0.4674   & 0.4674  & 0.4673  & 0.4673  \\
       &   1/3   & 0.2942   & 0.2942  & 0.2942  & 0.2942  \\
340    &   0     & 0.1178   & 0.1178  & 0.1177  & 0.1177  \\
       &   1/3   & 0.07413  & 0.07413 & 0.07411 & 0.07411 \\
370    &   0     & 0.04230  & 0.04229 & 0.04228 & 0.04228 \\
       &   1/3   & 0.02663  & 0.02662 & 0.02662 & 0.02662 \\
\noalign{\smallskip}
\end{tabular}
\end{table}

~\vfill\eject
\section*{Figure Captions}

Figure 1: Required destabilizing force profile of the radial component of the
          magnetic force density ${\rm f}_{{\rm K,r}}$ for inner heating (a). A fluid volume at
          the distance $r+\delta r$ (solid rectangle, (b)) experiences a larger force
          than a fluid volume at the distance $r$ (dashed rectangle, (b)). Moving the
          latter volume from $r$ to $r+\delta r$ (dot-dashed rectangle, (b)) results in
          an effective force which points in the direction of the displacement (dot-dashed
          arrow, (b)).

Figure 2: Region of potentially unstable and stable force profiles for a fixed outer
    temperature of $T_0=300$ K and $\chi_L=3$. The four different sets are 
    $\beta_1\!\!=\!\!0$, $N\!\!=\!\!1$ (long-dashed line), $\beta_1\!\!=\!\!0$, $N\!\!=\!\!0.7$
    (dot-dashed line), $\beta_1\!\!=\!\!1/3$, $N\!\!=\!\!1$ (solid line), and
    $\beta_1\!\!=\!\!1/3$, $N\!\!=\!0.7$ (dotted line), where
    the first and last one practically coincide.

Figure 3: Critical external induction $B_c$ versus inner temperature $T_1$ for a
    room temperature of $T_0\!=\!300$ K. For a horizontally infinitely extended layer,
    i.e. $R_{{\rm out}}\!=\!\infty$, the inclusion of a quadratic term in
    the susceptibility with $\beta_1\!\!=\!\!1/3$ (solid line) results in a lower threshold
    for the onset of convection than in the dilute case, $\beta_1\!=\!0$ (long-dashed line).
    Contrary for $R_{{\rm out}}\simeq 3.33$ cm, the critical induction for the dilute
    fluid (dot-dashed line) is lower than for the nondilute fluid (dotted line).
    The fluid parameters of the magnetic fluid EMG 901 and the size of the
    cylinder are given in the text.

Figure 4: Critical azimuthal wave number $l_c$ versus inner temperature $T_1$ for
    $T_0=300$ K. With decreasing temperature difference the wave number,
    i.e. the number of convection rolls, increases dramatically from $l_c\!=\!14$ for
    $\Delta T\!=\!70$ K to $l_c\!=\!315$ for $\Delta T\!=\!6$ ($\bullet$).

Figure 5: $f_{\chi_L}$, a measure for the strength of the magnetic force, versus
    the Langevin susceptibility $\chi_L$. The four different sets are 
    $\beta_1\!\!=\!\!0$, $N\!\!=\!\!1$ (long-dashed line), $\beta_1\!\!=\!\!1/3$, $N\!\!=\!\!1$
    (solid line), 
    $\beta_1\!\!=\!\!0$, $N(\gamma\!\!=\!\!0.15,\chi_{{\rm eff}})$ (dot-dashed line), and
    $\beta_1\!\!=\!\!1/3$, $N(\gamma\!\!=\!\!0.15,\chi_{{\rm eff}})$ (dotted line).
    The vertical solid line at $\chi_L=3$ is a guide for the eye.

\begin{thebibliography}{99}
\bibitem{rosensweig}
R. E. Rosensweig, {\it Ferrohydrodynamics} (Cambridge University Press,
Cambridge, 1985).

\bibitem{finlayson}
B. A. Finlayson, ``Convective instability of ferromagnetic fluids'', J. Fluid Mech. {\bf 40}, 753 (1970).

\bibitem{schwab}
L. Schwab, U. Hildebrandt, and K. Stierstadt, ``Magnetic {B{\' e}nard} convection'', J. Magn. Magn. Mat. {\bf 39},
113 (1983); L. Schwab and K. Stierstadt, ``Field-induced wavevector-selection by magnetic
{B{\' e}nard}-convection'', ibid. {\bf 65}, 315 (1987); 
L. Schwab,  ``Thermal convection in ferrofluids under a free surface'', ibid. {\bf 85}, 199 (1990).

\bibitem{auernhammer00}
G. K. Auernhammer and H. R. Brand, ``Thermal convection in a rotating layer of magnetic fluid'',
Eur. Phys. J. B {\bf 16}, 157 (2000).

\bibitem{huang}
J. Huang, B. F. Edwards, and D. D. Gray, ``Magnetic control of convection in nonconducting paramagnetic fluids'',
Phys. Rev. E {\bf 57}, R29 (1998); ``Thermoconvective instability of paramagnetic fluids in a nonuniform magnetic
field'', ibid. 5564.

\bibitem{morimoto98}
H. Morimoto, T. Maekawa, and M. Ishikawa, ``Linear stability analysis of magnetic {Rayleigh-B{\' e}nard}
convection'', Adv. Space Res. {\bf 22}, 1271 (1998).

\bibitem{zebib96}
A. Zebib, ``Thermal convection in a magnetic fluid'', J. Fluid Mech. {\bf 321}, 121 (1996).

\bibitem{odenbach93}
S. Odenbach, ``Drop tower experiments on thermomagnetic convection'', Microgravity sci. technol. {\bf VI}, 161
(1993).

\bibitem{du98}
T. Du and W. Luo, ``Nonlinear optical effects in ferrofluids induced by temperature and concentration cross
coupling'', Appl. Phys. Lett. {\bf 72}, 272 (1998).

\bibitem{luo99}
W. Luo, T. Du, and J. Huang, ``Field-induced instabilities in a magnetic fluid'', Phys. Rev. Lett. {\bf 82},
4134 (1999).

\bibitem{shliomis00}
M. I. Shliomis and M. Souhar, ``Self-oscillatory convection caused by the {Soret} effect'', Europhys. Lett.
{\bf 49}, 55 (2000).

\bibitem{schaertl99}
W. Schaertl and C. Roos, ``Convection and thermodiffusion of colloidal gold tracers by light scattering'',
Phys. Rev. E {\bf 60}, 2020 (1999).

\bibitem{spill00}
R. Spill, W. K\"ohler, G. Lindenblatt, and W. Schaertl, ``Thermal diffusion and {Soret} feedback of gold-doped
polyorganosiloxane nanospheres in toluene'', Phys. Rev. E {\bf 62},
8361 (2000).

\bibitem{shliomis_01}
M. I. Shliomis, ``Comment on `Novel convective instabilities in a magnetic fluid'~'', Phys. Rev. Lett. {\bf 87},
059801 (2001).

\bibitem{lu01}
W. Luo, T. Du, and J. Huang, ``{Luo, Du, and Huang} reply'', Phys. Rev. Lett. {\bf 87}, 059802 (2001).

\bibitem{onsager36}
L. Onsager, ``Electric moments of molecules in liquids'', J. Am. Chem. Soc. {\bf 58}, 1486 (1936).

\bibitem{mean_spheric}
M. S. Wertheim, ``Exact solution of the mean spherical model for fluids of hard spheres with permanent electric
dipole moments'', J. Chem. Phys. {\bf 55}, 4291 (1971);
K. I. Morozov and A. V. Lebedev, ``The effect of magneto-dipole interactions on the magnetization curves of
ferrocolloids'', J. Magn. Magn. Mat. {\bf 85}, 51 (1990).

\bibitem{buyevich92}
Y. A. Buyevich and A. O. Ivanov, ``Equilibrium properties of ferrocolloids'', Physica A {\bf 190}, 276 (1992).

\bibitem{chen91}
D. Chen, J. A. Brug, and R. B. Goldfarb, ``Demagnetizing factors for cylinders'', IEEE Trans. Magnetics {\bf 27},
3601 (1991).

\bibitem{lange01_kelvin}
A. Lange, ``Kelvin force in a Layer of Magnetic Fluid'', to appear in J. Magn. Magn. Mat., (2002).

\bibitem{voelker}
T. V\"olker, E. Blums, and S. Odenbach, ``Thermodiffusive processes in ferrofluids'', Magnetohydrodynamics
{\bf 37}, 274 (2001).

\bibitem{chandrasekhar}
S. Chandrasekhar, {\it Hydrodynamic and hydromagnetic stability}, (Dover
Publications, New York, 1981), App. V.

\bibitem{comment}
The thermal diffusivity $\kappa =\lambda/(\rho c_p)$ can
only be estimated since the producer does not report
the heat capacity $c_p$ and the thermal conductivity $\lambda$ of the fluid EMG 901.

\bibitem{luo99_jmmm}
W. Luo, T. Du, and J. Huang, ``Field-induced instabilities in a magnetic fluid'', J. Magn. Magn. Mat. {\bf 201},
88 (1999).
\end{thebibliography}
\end{document}